# Chain Connectivity and Conformational Variability of Polymers: Clues to an Adequate Thermodynamic Description of their Solutions

## III: Modeling of Phase Diagrams

Sergej Stryuk and Bernhard A. Wolf*

Institut für Physikalische Chemie und Materialwissenschaftliches Forschungszentrum
der Johannes Gutenberg-Universität Mainz,
Jakob Welder-Weg 13, D-55099 Mainz, Germany
FAX: +(0)6131-39-24640
Electronic mail: bernhard.wolf@uni-mainz.de

**Full Paper:** A simple expression for the composition dependence of the Flory-Huggins interaction parameter of polymer/solvent systems reported earlier is used to model the demixing of polymer solutions into two liquid phases. To this end the system specific parameters $z$ and $n$ of that approach are calculated as a function of temperature using the thermodynamic expressions resulting for the critical conditions on one side and from experimentally determined critical data for polymers of different molar mass on the other side. By means of data reported for the system cyclohexane/polystyrene it is demonstrated that binodal and spinodal lines are very accurately modeled at low temperatures (UCSTs) and at high temperatures (LCSTs). The parameters obtained from the demixing behavior match well with that calculated from the composition dependence of the vapor pressure at temperatures where the components are completely miscible. Information on the phase separation of the system *trans*-decalin/polystyrene for different molecular weights and at different elevated pressures is used to show that the approach is also apt to model pressure influences. The thus obtained $z\,(T;p)$ and $n\,(T;p)$ enable the prediction of the (endothermal) theta temperature of the system as a function of pressure in quantitative agreement with the data directly obtained from light scattering measurements with dilute solutions.

## Introduction

It is a matter of common knowledge that the Flory-Huggins theory suffers from numerous deficiencies. The most outstanding shortcoming consists in the fact that it does not account for the inhomogeneous distribution of the polymer segments in dilute solutions. Furthermore it cannot explain the experimentally observed dependence of interaction parameters on chain length even for very large chain overlap close to the pure polymer melt. In view of this situation several very sophisticated attempts have been made to improve the thermodynamic description of polymer containing mixtures[1]. Most of them are rather complicated and not very handy for practical purposes, moreover they still require adjustable parameters for a quantitative description of experimental results in many cases. For this reason we have recently established an approach that makes explicit allowance for chain connectivity (i.e. inhomogeneous space filling at high dilution) and conformational relaxation (i.e. influences of chain length at high polymer concentrations). This concept has turned out to be able to explain hitherto unconceivable finding concerning the dependence of second osmotic virial coefficients on the molar mass of the polymer[2] and to describe the very diverse experimental data concerning the composition dependence of the Flory-Huggins interaction parameter $c$ (like the occurrence of minima)[3]. In this contribution we check to which extent this approach can model demixing and whether the information obtained from homogenous mixtures and from phase diagrams match.

For that purpose we have taken experimental demixing data published for solutions of polystyrene (PS) in either cyclohexane (CH) or *trans*-decalin (TD) and used the theoretical equations worked out in the preceding publications[2,3] to model the phase diagrams. The information concerning the homogeneous state is only available for CH/PS and stems from vapor pressures measured at two temperatures slightly higher than the (endothermal) theta temperature of the system[4].

## Theoretical part

### *Composition dependent interaction parameters*

The present approach[3] requires four parameters for an analytical representation of experimental data, out of which one can in most cases be treated as a known constant and two of the remaining three parameters are interdependent. Because of the fact that the interaction parameters normally depend on composition, different equations are obtained[5] for the differential interaction parameters based on the chemical potentials of the components and for the integral interaction parameter $g$.

The segment molar Gibbs energy of mixing is generally written as

$$\frac{\Delta \bar{\bar{G}}}{RT} = (1-j)\ln(1-j) + \frac{j}{N}\ln j + g(1-j)j \qquad (1)$$

where $j$ is the volume fraction of the polymer and $N$ the number of segments, defined as its molar volume divided by the molar volume of the solvent. The present approach yields the following expression[3] for $g$

$$g = \frac{a}{(1-n)(1-nj)} - z\left(1+(1-l)j\right) \qquad (2)$$

where we can for many purposes in good approximation set $l$ equal to 0.5. The parameter $a$ quantifies the thermodynamic effect of opening a contact between polymer segments by insertion of a solvent molecule at infinite dilution without changing the conformation of the polymer chain. This first step of the mixing process does in general not describe the total effect. It is only via conformational relaxation, quantified by the parameter $z$, that the equilibria are reached. The rearrangement of segments in response to a change in the immediate environment is only absent under theta conditions when $z$ becomes zero and $a$ assumes the value of 0.5. Finally, $n$ accounts primarily for the change in the deviation of the entropy of mixing from combinatorial behavior with composition. This discrepancy is larger at low polymer concentrations (inhomogeneous space filling of polymer segments) than at high polymer concentrations (where the reservoir of pure solvent has already been emptied).

For the modeling of phase diagrams Eq. (2) suffices because we use the direct minimi-zation of the Gibbs energy[6] for that purpose. However, for the evaluation of experimental information on the chemical potential of the solvent (e.g. via vapor pressures) as a function of composition, the corresponding differential expression is more adequate. According to the present approach it reads

$$c = \frac{a}{(1-n\,j)^2} - z\,l\left(1+2\left(\frac{1}{l}-1\right)j\right) \qquad (3)$$

From the fact that both steps of mixing are governed by the thermodynamic quality of a given solvent it becomes immediately obvious that $a$ and $z$ cannot be independent of each other. Indeed the analysis of extensive experimental material concerning the chain length dependence of the second osmotic virial coefficient[2] and on the composition dependence of the Flory-Huggins interaction parameter has demonstrated that the following simple relation holds true for common linear vinyl polymers[3]

$$a = 0.5 + \mathrm{D}\,z\,l \qquad (4)$$

where the constant D is less than unity. To which extent the above relation remains valid for non-vinyl polymers or macromolecules of non-linear architecture and whether a linear interrelation suffices to cover the entire range from marginal to extremely good solvents requires further investigation.

*Critical conditions and modeling*

As a consequence of the requirement that the 2nd and 3rd derivatives of the Gibbs energy of mixing be zero one obtains the following relations from Eqs. (1) and (2):

$$\frac{1}{1-j_c} + \frac{1}{N j_c} + \frac{2a}{(n j_c - 1)^3} + 2z\left[l - 3j_c(l-1)\right] = 0 \qquad (5)$$

plus

$$\frac{1}{(j_c-1)^2} - \frac{1}{N j_c^{\,2}} - \frac{6an}{(n j_c - 1)^4} + 6z(1-l) = 0 \qquad (6)$$

Inserting $z$ from Eq. (6) into Eq. (5) and substituting for $a$ according to relation (4) yields

$$\frac{1}{1-j_c} + \frac{1}{N j_c} + \frac{1}{(n j_c - 1)^3} + \frac{\left(\frac{1}{(j_c-1)^2} - \frac{1}{N j_c^{\,2}} - \frac{6n}{(n j_c - 1)^4}\right)}{\left(\frac{6\mathrm{D}n\,l}{(n j_c - 1)^4} - 6(1-l)\right)} * \left(\frac{2\mathrm{D}l}{(n j_c - 1)^3} + (2 - 6j_c + \frac{6j_c}{l})\right) = 0 \qquad (7)$$

where $j_c$ is the critical volume fraction of the polymer.

In order to calculate the system specific parameter $n$ from experimentally determined critical data, one inserts the numerical values for D and $l$ and solves Eq. (7) by means of an iterative procedure, where only one of the solutions is physically meaningful. D represents a constant for a given system (here D=0.59 according to the adjustment) and $l$ can either be calculated from the parameters of the Kuhn-Mark-Houwink equation as described[2] or can be set in very good approximation equal to 0.5 for most systems. After that $z$ is calculated according to Eq. (6) as

$$z = \frac{\left[\frac{1}{(j_c-1)^2} - \frac{1}{N j_c^{\,2}} - \frac{3n}{(n j_c - 1)^4}\right]}{\left(\frac{6\mathrm{D}n\,l}{(n j_c - 1)^4} - 6(1-l)\right)} \qquad (8)$$

The thus determined system specific parameters $z$ and $n$ do naturally refer to one temperature only, namely to the critical temperature $T_c$ for the solutions of a given polymer consisting of $N$ segments. To enable the modeling of the complete phase diagrams it is therefore necessary to repeat the procedure and to evaluate critical data for several polymers differing sufficiently in molar mass. In this manner one obtains access to the temperature dependence of $z$ and $n$.

On the basis of this information it is possible to calculate binodal and spinodal lines according to the normal thermodynamic techniques. In this work we have again minimized the Gibbs energy of the system directly; this procedure[6] does not require the calculation of chemical potentials and is particularly apt for complicated or non-analytical expressions for the Gibbs energy as a function of composition and temperature.

*Vapor pressures and interaction parameters*

The vapor pressure of the solvent above a polymer solution ($p_1$), as compared with the vapor pressure of the pure solvent ($p_{1,o}$), yields access to the activity $a$ of this component in the mixture, i.e. on the corresponding (differential) interaction parameter $c$ according to

$$\ln\frac{a_1}{a_{1,o}} = \ln(1-j) + \left(1-\frac{1}{N}\right) j + c\, j^2 \qquad (9)$$

This means that experimental information on the composition dependence of $p_1$ yields the system specific parameters of the present approach in case of completely miscible components. For the determination of $z$ and $n$ according to Eq. (3) we have again used the interrelation Eq. (4) with D=0.59 and approximated $l$ by 0.5. The fitting can be easily done by means of conventional programs.

## Evaluation of experimental data

*Cyclohexane/polystyrene*

The system cyclohexane/polystyrene (CH/PS) is one of the best studied in the field of thermodynamics and tested first with respect to the possibilities to model phase diagrams. Extensive data for solutions of polymers differing widely in their molar masses at normal pressure (demixing upon cooling, UCSTs near room temperature) and under the equilibrium vapor pressure of the solvent (demixing upon heating, LCST up to 240 °C) have been published[7]. These results are shown in Fig. 1 together with the binodal and spinodal curves calculated according to the present approach.

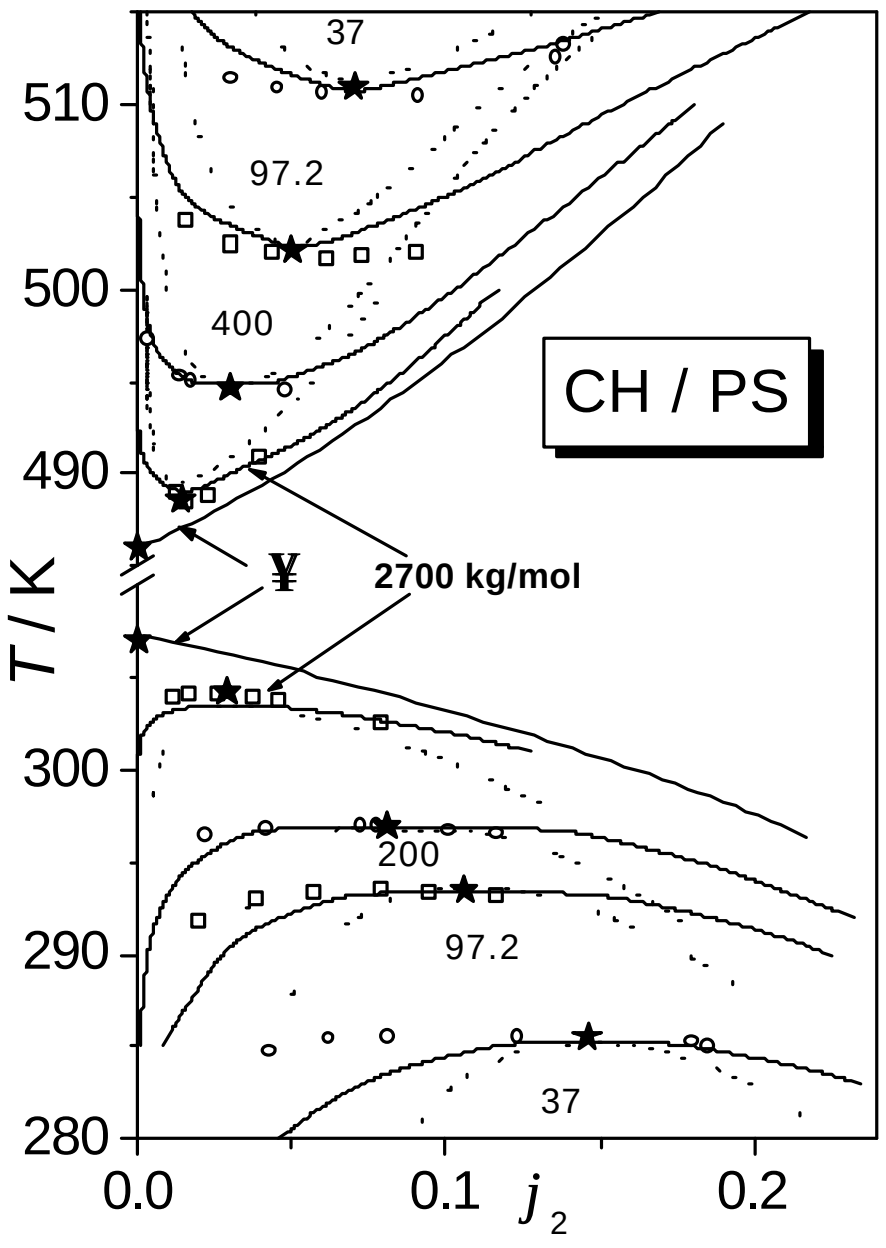

Fig. 1: Phase diagram (demixing into two liquid phases) of the system CH/PS for the indicated molar masses of the polymer (kg/mol). Cloud points (under the equilibrium vapor pressure of the solvent at the high temperatures) are taken from the literature[7]; binodals (full lines) and spinodals (dotted lines) were calculated as described in the text by means of the temperature dependent parameters $z$ and $n$ (cf. Fig. 2). The critical points are represented by full stars.

The variation of the conformational response *z* and of the parameter *n* (as calculated from the critical data shown in Fig. 1 according to the procedure described in last section) are presented in Fig. 2. Negative *z* values correspond to worse than theta conditions. This parameter becomes zero at the UCST and at the LCST for infinitely high *M* values, which are normally identical with the corresponding endothermal and exothermal theta temperature, respectively.

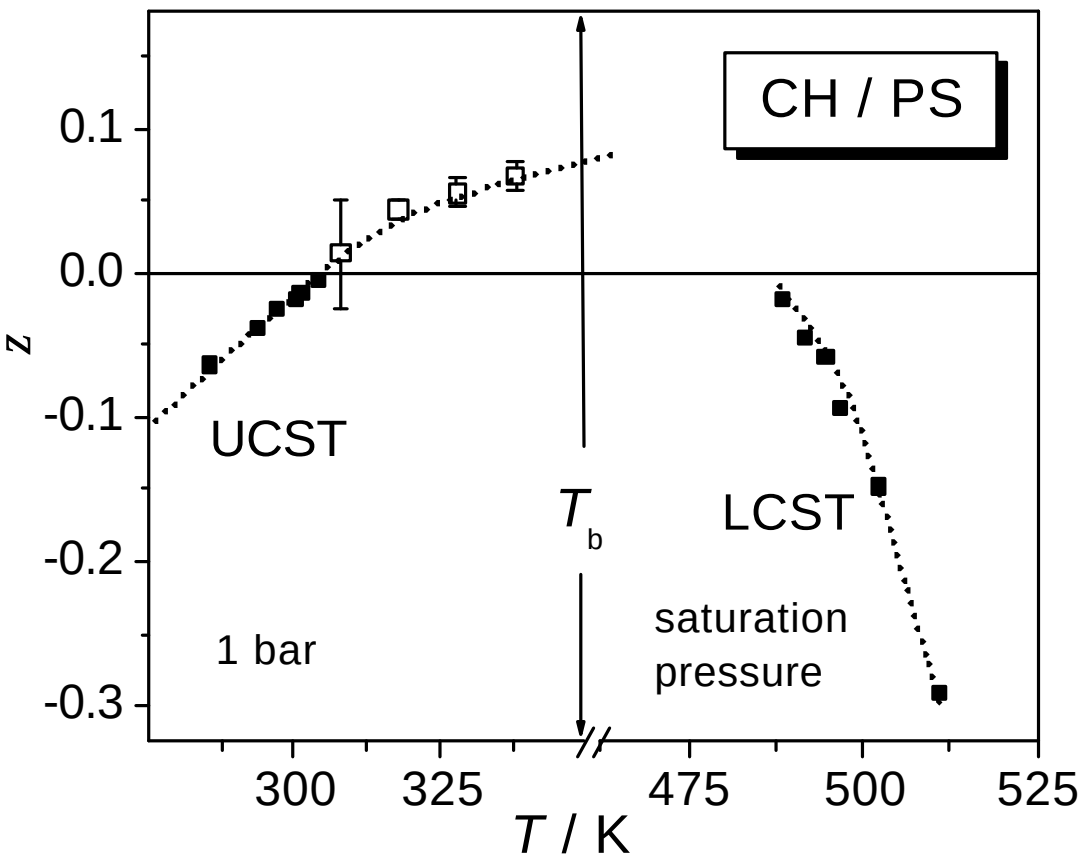

Part a

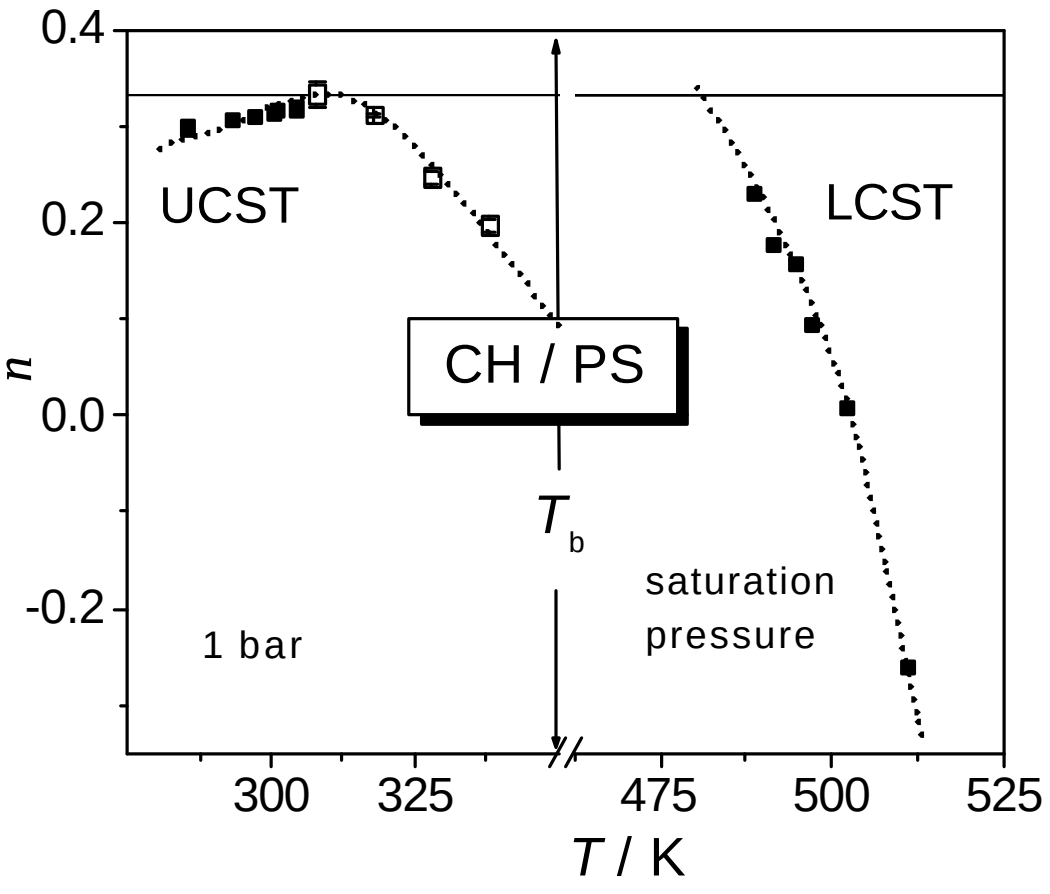

Part b

Fig. 2: Temperature dependence of the conformational response *z* (part a) and of the parameter *n* (part b) as obtained from the evaluation of the critical data shown in Fig. 1 (full symbols). Also incorporated are the results from vapor pressure data[4,5] (open symbols) (cf. Fig. 3). Up to the boiling point of the solvent, $T_b$, the information refers to 1 bar. For the endothermal theta temperature of the system the Polymer Handbook[8] states values from 307.1 to 308.1 K.

Fig. 2 aggregates the information from two sources: Critical data of polymers differing in chain length and vapor pressures of homogeneous solutions as a function of composition. Fig. 3 gives an example for the composition dependence of the interaction parameters resulting from solvent activities[4,5] and a temperature which is slightly higher than the (endothermal) theta temperature of the system. The *z* values plotted in Fig. 2a differ to some extent from that reported earlier[3] due to a modified evaluation procedure of the primary data. In the previous work we have assumed that $c_o$, the Flory-Huggins interaction parameter in the limit of infinite dilution (very accurately measured via the second osmotic virial coefficients) can be used to eliminate one of the system specific parameters. Here we treat $c_o$ like all other *c* values and use Eq. (4) to reduce the number of adjustable parameters. The comparison of the results demonstrates the superiority of the latter procedure: Except for the immediate vicinity of the theta temperature the uncertainty of the parameters is considerably less.

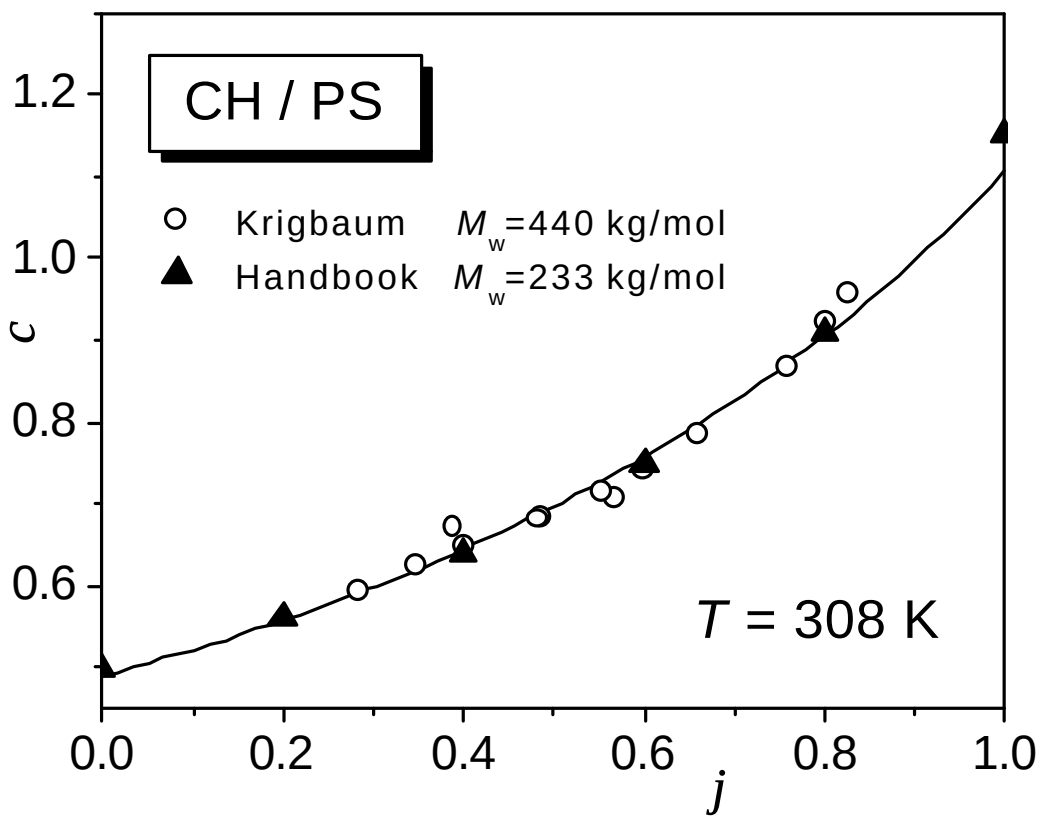

Fig. 3: Data points: Flory-Huggins interaction parameter *c* calculated from vapor pressure data[4,5] for 308 K. The curve is modeled according to Eq. (3). Due to the proximity of *T* to the theta temperature of the system, the differences resulting from dissimilar chain lengths remain negligible[9].

Despite the fact that the data published in the literature for the system CH/PS cover a temperature range of more than 200 °C it is difficult to draw firm conclusions concerning the typical $T$-dependence of $z$ and $n$ within the entire interval from endothermal to exothermal conditions. The reason lies in the fact that all information concerning the lower temperatures refers to 1 bar and consequently ends at the boiling point $T_b$ of the solvent indicated in Fig. 2. At high temperatures, on the other hand, the pressure rises constantly from 16 bar (the vapor pressure of the solution of critical composition for the largest molecular weight PS) up to 23 bar. In order to obtain consistent information over the entire temperature range and to be capable of splitting the parameters into enthalpy and entropy contributions, measurements at $p \geq$ 23 bar would be required. Some conclusions concerning the situation in the vicinity of UCSTs can, however, be drawn: First of all, the $z$ values obtained from liquid/liquid demixing and that resulting from vapor pressure are consistent. They evidence that the improvement of solvent quality with rising temperature is in this temperature range primarily due to conformational relaxation.

According to the present findings $n$ passes a maximum at the theta temperature. This result can be tentatively rationalized by the following consideration: At $T = Q$ the polymer coils assume their unperturbed dimensions, the second osmotic virial coefficient becomes zero ($z$ = 0 and $a = c_o = 0.5$) and mixing is random. Due to a compensation of different effects there is no preference of certain types of contacts between polymer segments and solvent molecules. Despite this situation the deviation from combinatorial behavior increases as the polymer concentration rises because of the growing consumption of the reservoir of pure solvent (existing outside the domains occupied by polymer coils). Within the scope of the present approach this increasing deviation from combinatorial behavior is quantified by the parameter $n_\Theta$. As the solvent becomes better than a theta solvent, contacts between polymer segments and solvent molecules are preferred over like contacts in contrast to worse than theta solvents, where the reverse is true. In both cases the establishment of quasi-chemical equilibria offers a route for a further reduction of the Gibbs energy[10] via the best compromise between the gain associated with the formation of the preferred contacts and the loss resulting from non-random mixing. This additional contribution depends on composition and reduces the Flory-Huggins parameter. In the light of the present approach it is the reason why $n$ passes a maximum at the theta temperature in the absence of preferred contacts.

Before we switch to the system TD/PS it is worthwhile to analyze the critical data in more detail. According to the original Flory-Huggins theory[11] the critical volume fraction of the polymer should depend on $N^{-0.5}$. The following graph shows how $j_c$ varies with the number of polymer segments in the present case.

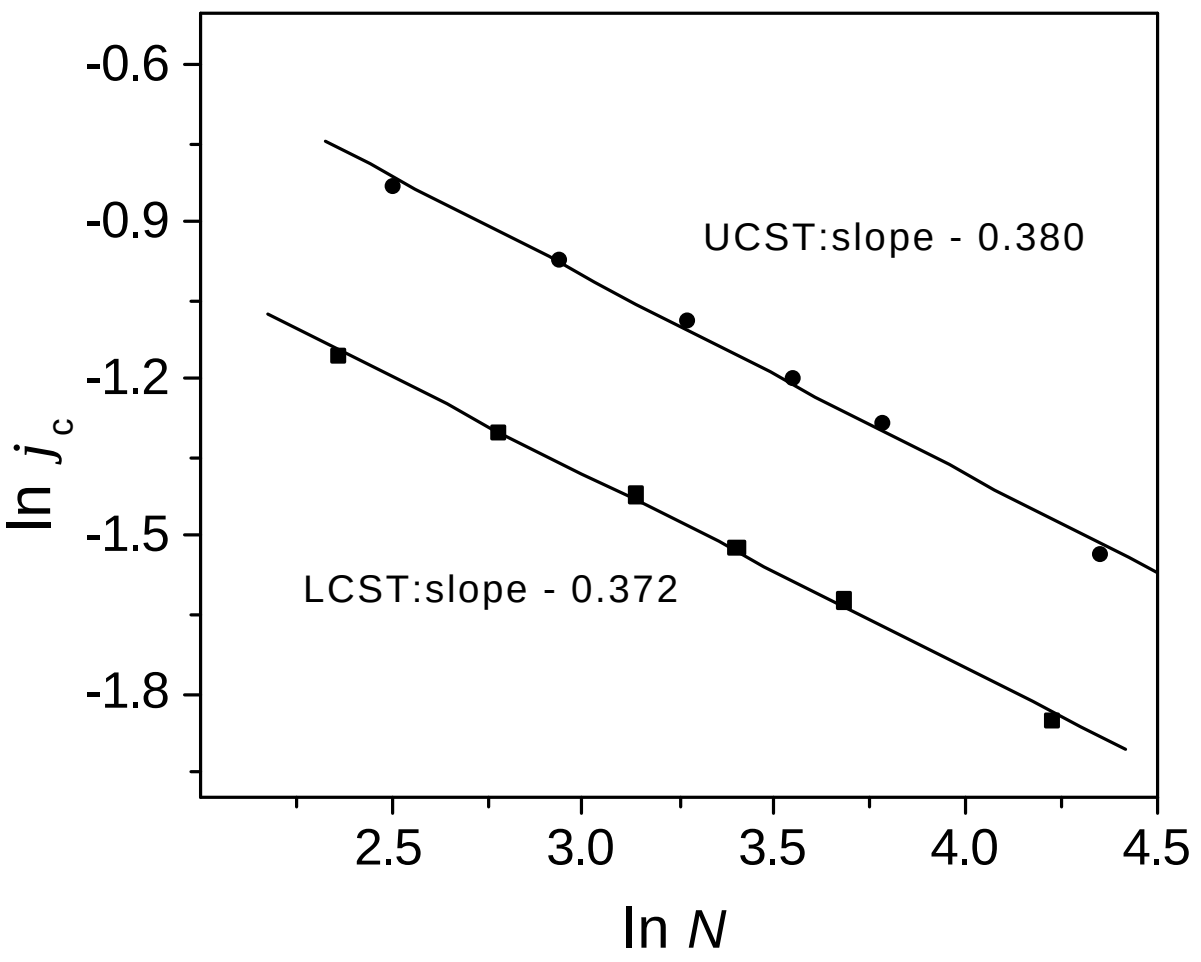

Fig. 4: Double logarithmic plot of the dependence of the critical polymer concentration $j_c$ on the number of polymer segments $N$ for the system CH/PS.

The reduction of $j_c$ upon an augmentation of $N$ is considerable less than postulated by the Flory-Huggins theory as demonstrated by the slopes of the lines in Fig. 4. Strictly speaking the evaluation is only correct for the UCSTs where all data refer to 1 bar; in the case of the LCSTs the pressure increases as the chains become shorter. Another interesting feature consists in the actual values of the critical volume fractions for a given polymer sample, which are considerably lower for the LCSTs than for the UCSTs. The obvious interpretation of these differences in terms of the much larger free volume in the former case does not hold true, because the observed discrepancies are by far too large. Presently the reasons for that feature are unclear.

A closer inspection of experimental information on $j_c$ ($N$) reveals that the following relation is obeyed by most systems.

$$j_c = A\, N^{-B} \quad \text{with } B < 0.5 \qquad (10)$$

where A and B are constants for a given system. Under these conditions it is possible to calculate the value, $n$ must assume at the critical temperature in the limit of infinite molar mass of the polymer (theta temperature). Substituting for $j_c$ from Eq. (10) in the second term of Eq. (6) and knowing that the conformational response $z$ becomes zero under theta condition leads to the following expressions

$$v_\Theta = \frac{1}{3} \quad \text{where} \quad z_\Theta = 0 \qquad (11)$$

which are within experimental error well fulfilled by the present system, as can be seen from Fig. 2. At the Θ temperatures (endothermal and exothermal) read from *part a* of this graph ($z_\Theta = 0$) the parameter $n$ depicted *in part b* assumes the value 1/3.

*Trans-decalin/polystyrene*

In order to check whether the present approach can also describe the influences of pressure on demixing, we have modeled published data[12] for the system *trans*-decalin/polystyrene (TD/PS). The phase diagrams for normal pressure and three molar masses of the polymer are shown in Fig. 5. The (limited) experimental information on $j_c$ ($N$) yields B=0.25 (Eq. (10)).

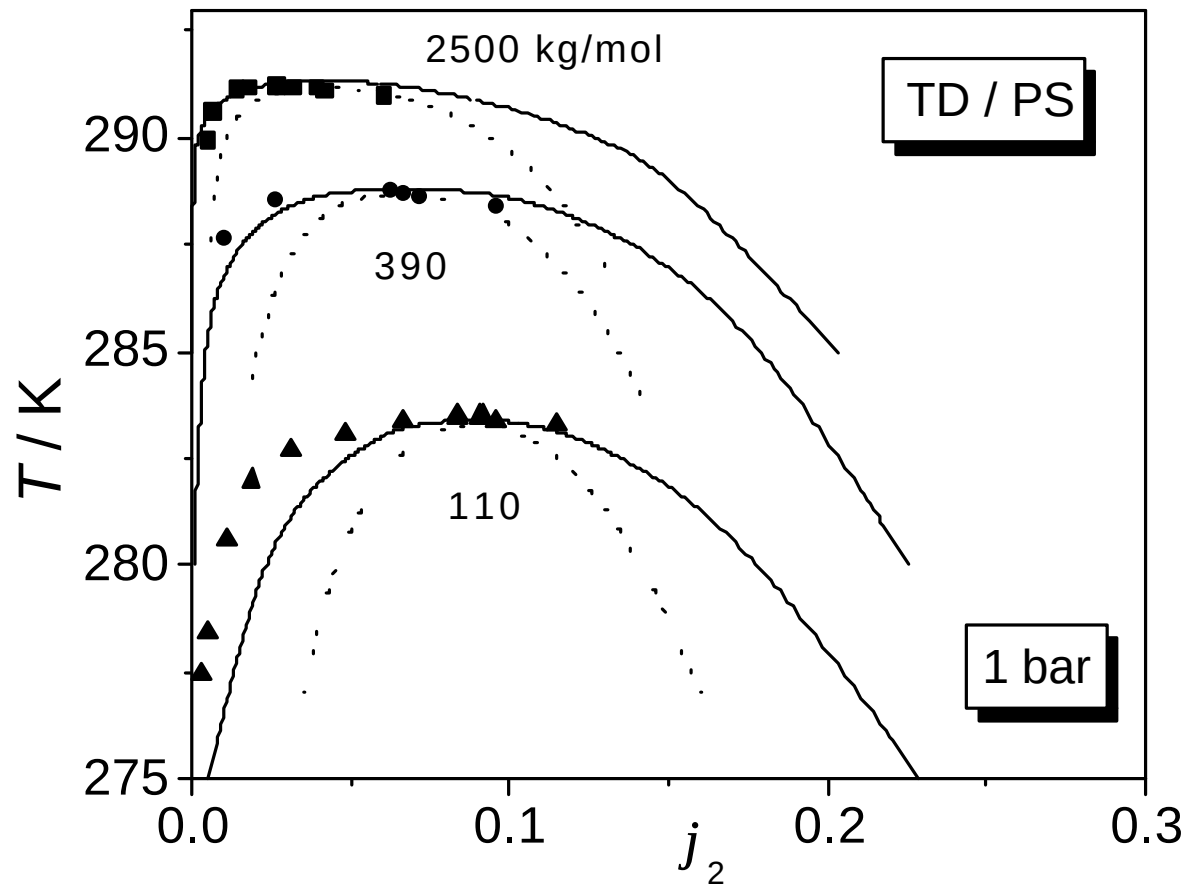

Fig. 5: Phase diagram of the system TD/PS for the indicated molar masses of the polymer (kg/mol). Cloud points are taken from the literature[12] ; binodals (full lines) and spinodals (dotted lines) were calculated as described in the text.

Modeling of the above phase diagrams results in the parameters $z$ and $n$ and their variation with temperature shown in Fig. 6. The theta temperature read from the condition (11) agrees well with that reported in the literature[12].

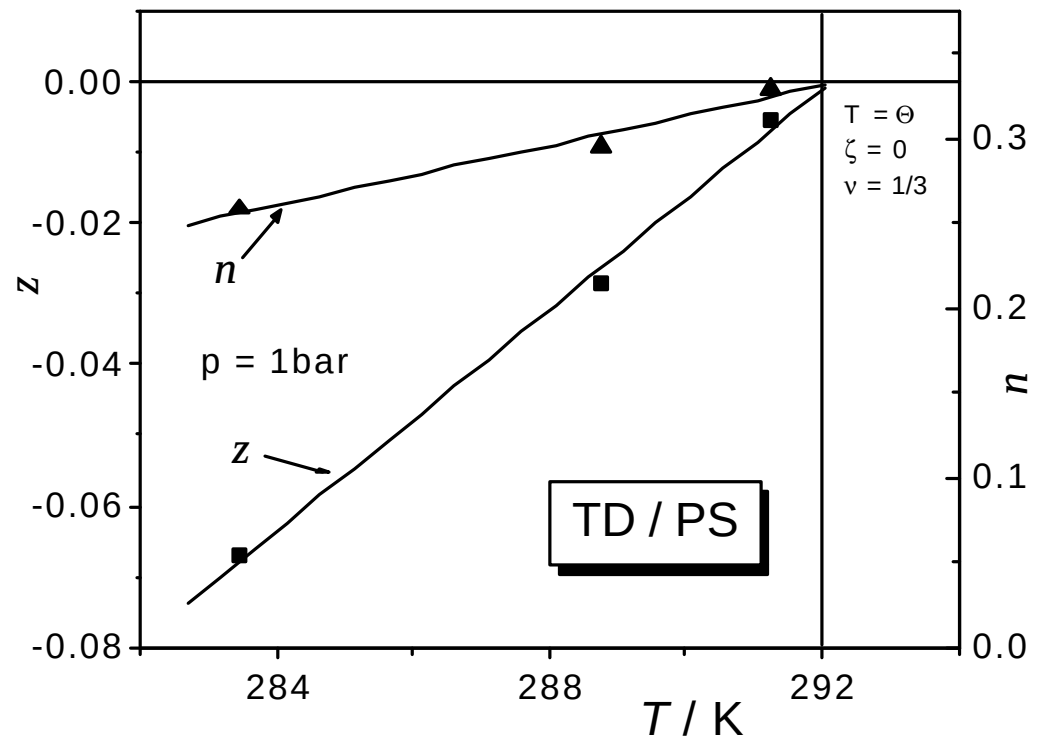

Fig. 6: Temperature dependence of the conformational response $z$ and of the parameter $n$ resulting from the evaluation of the critical data read from Fig. 5.

How the system specific parameters obtained from the demixing data published for elevated pressures[12] vary with $T$ is shown in Fig. 7.

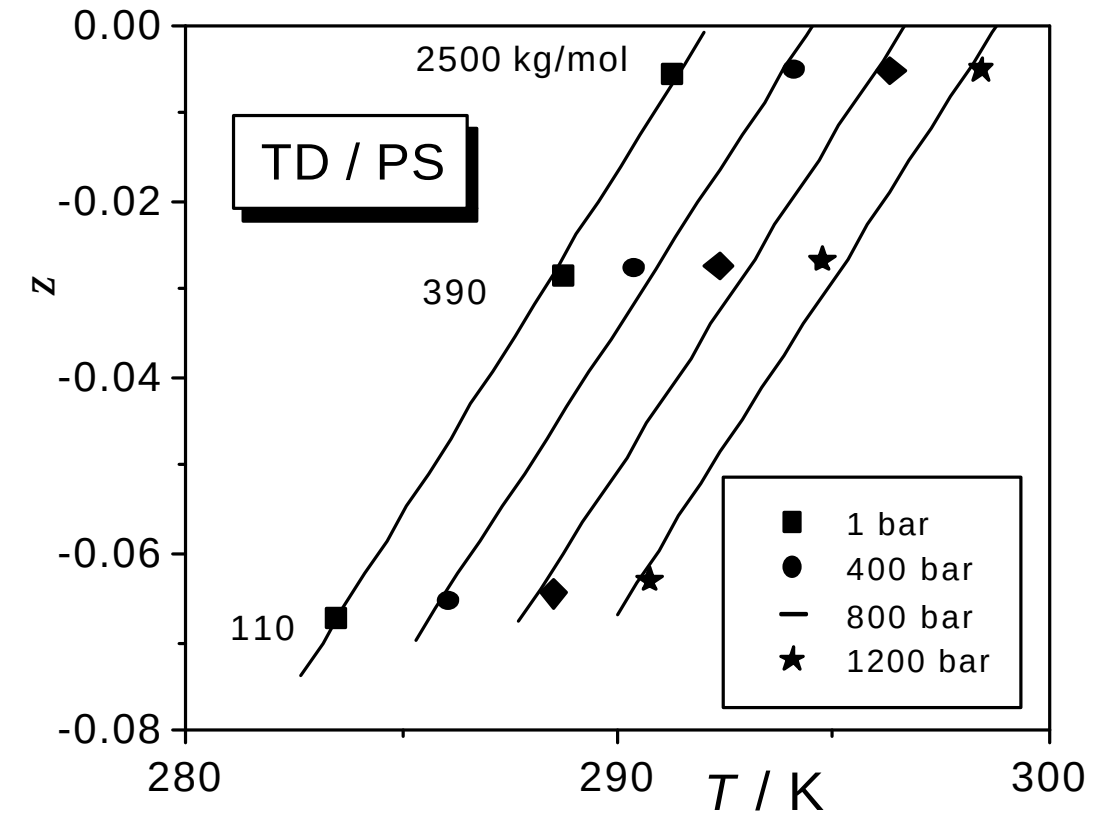

Part a

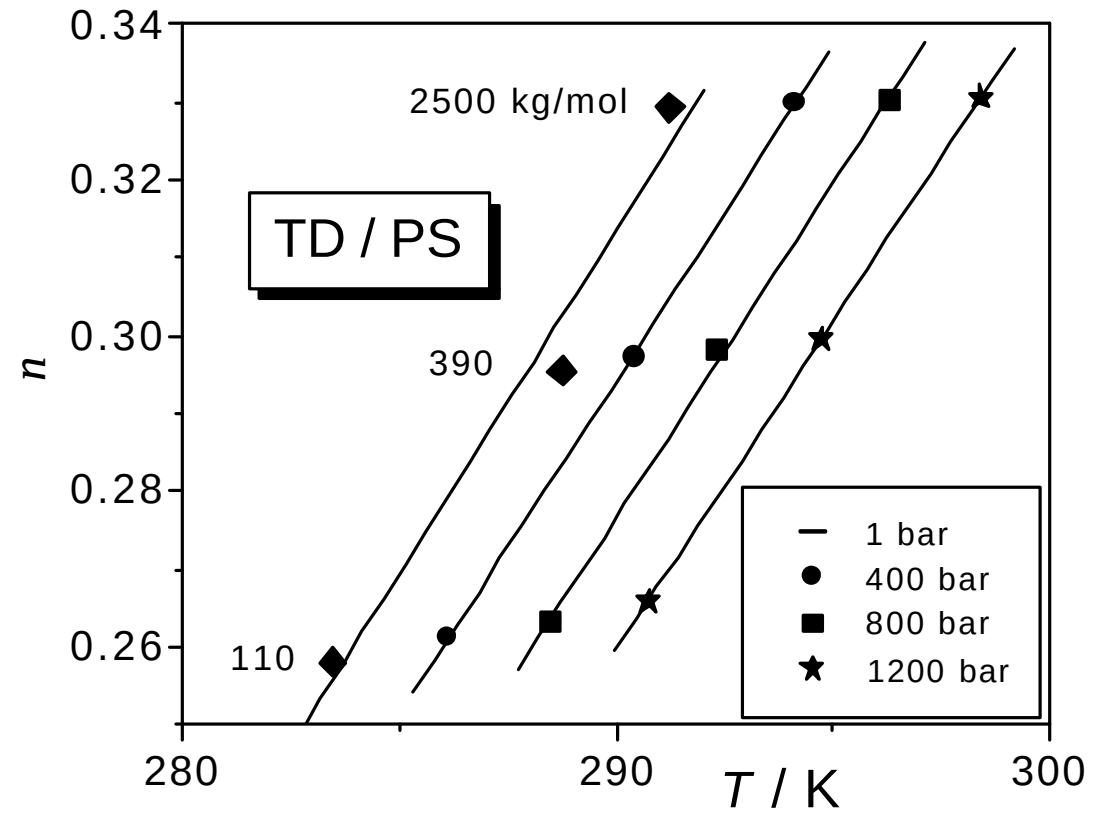

Part b

Fig. 7: Temperature dependence of the parameters $z$ (part a) and $n$ (part b) at the indicated constant pressures as calculated from critical lines[12] of the system TD/PS.

Keeping in mind that $z$ is much more important for the thermodynamic quality of the solvent than $n$ (which does for instance in the limit of high dilution not at all contribute to $c$) it becomes immediately obvious from part a of Fig. 7 that the mixing tendency decreases with falling temperature and rising pressure (reduction of $z$).

How the phase diagrams calculated according to the present approach for different constant pressures look like in detail is exemplified in Fig. 8 for the middle molecular weight sample of PS.

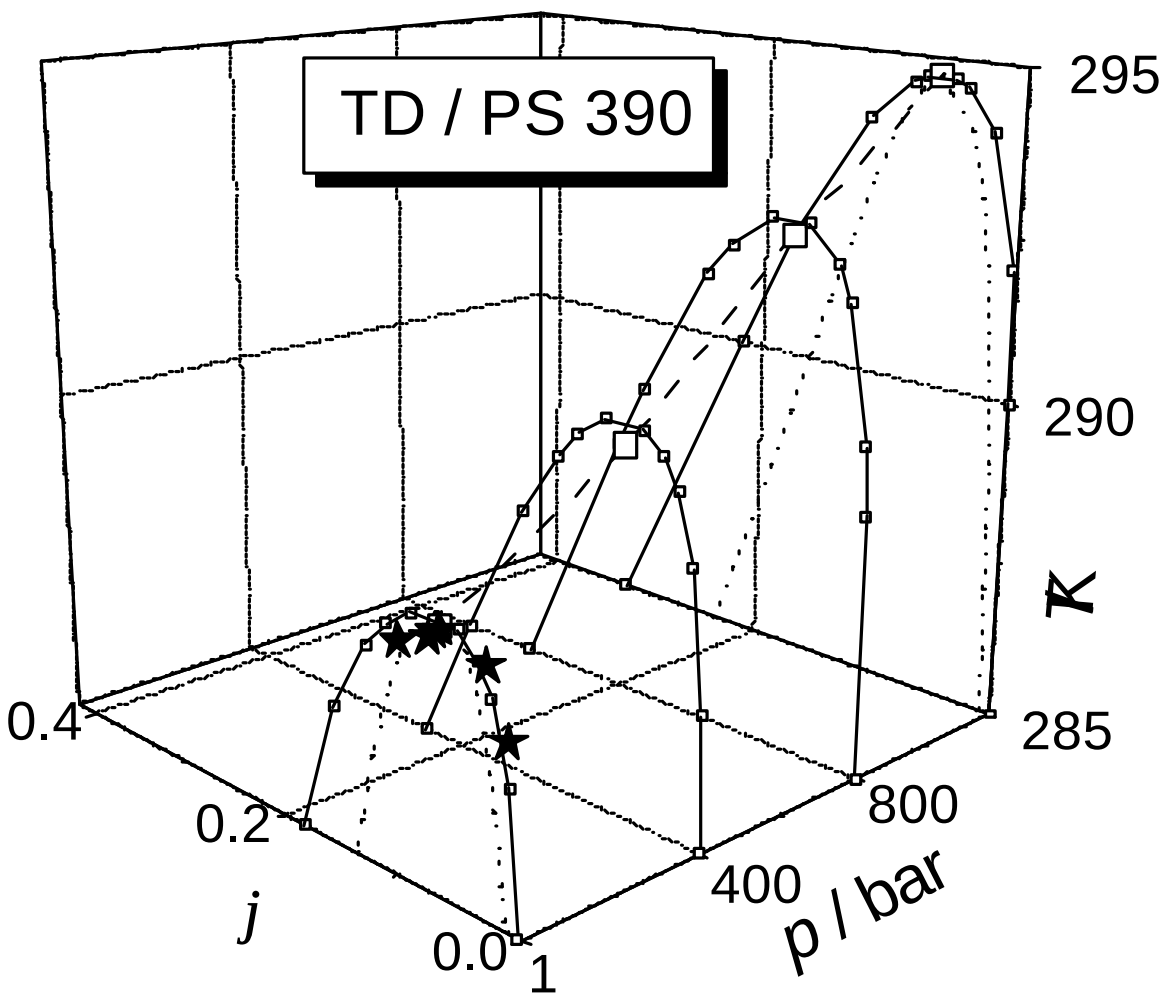

Fig. 8: Binodal curves (full lines) and some spinodal curves (dotted lines) calculated for the system TD/PS 390 by means of the parameters shown in Fig. 7. Also indicated are the cloud points (stars) measured under atmospheric pressure and $j_c\,(T, p)$ (broken line).

The temperature and pressure dependencies of $z$ and $n$ (cf. Fig. 7) do not only allow the modeling of phase diagrams, but also give access to the variation of the theta temperature with pressure via the conditions formulated in Eq. (11). For TD/PS direct information on the endothermal theta temperature $Q_+$ and its pressure dependence is already available from light scattering measurements in the composition range of pair interaction[12], where the theta conditions are evidenced by vanishing second osmotic virial coefficients $A_2$. Fig. 9 documents that the two sets of data agree quantitatively within a reasonable experimental error of ± 1 °C (interpolation in case of $A_2$ and extrapolation of either $z$ or $n$).

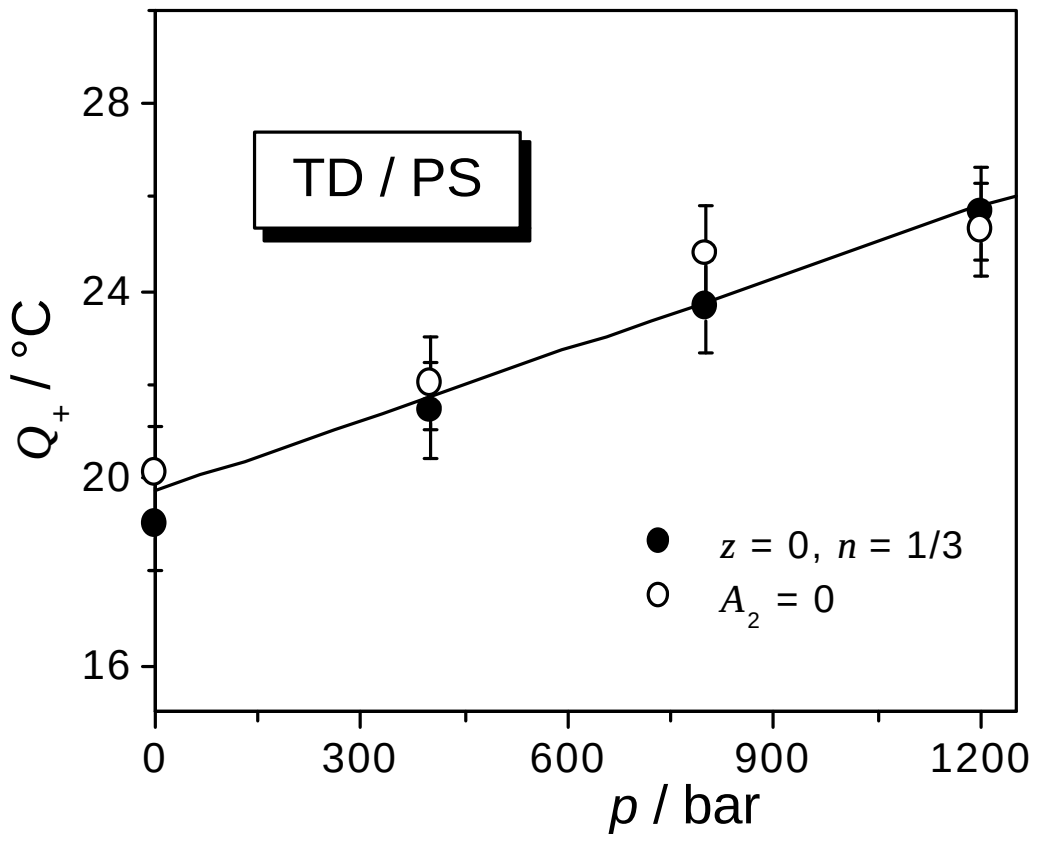

Fig. 9: Comparison of the pressure dependence of the endothermal theta temperature $Q_+$ as obtained from light scattering at low polymer concentration[12] and from the extrapolation of the system specific parameters of the present approach to their theta values.

## Conclusions

The present results demonstrate that phase diagrams of polymer/solvent systems can be modeled very accurately by the current approach, which accounts explicitly for chain connectivity and for the conformational variability of polymers. Subject to the condition that reliable critical data for solutions of sufficiently different molar masses of the polymer in a given solvent are available it is possible to calculate binodal lines and spinodal lines in good agreement with actual measurements. An observation that deserves further attention is the systematic deterioration of the agreement between experiment and theory as the molar mass of the polymer decreases and becomes on the order of 100 kg/mol or less. This feature, which can be clearly seen from Fig. 1 and from Fig. 5, is probably due to that fact that further members of the series expansion[2] of the logarithm of (1-$F_o$) (where $F_o$ is the volume fraction of segments in an isolated polymer coil) must be taken into account to describe the behavior of very short polymer chains (characterized by larger $F_o$ values).

A substantial argument in favor of the soundness of the physical considerations underlying the new approach consists in the fact that the parameters obtained from three different experimental sources agree quantitatively. The congruence of $z$ and $n$ values resulting from the molecular weight dependence of the second osmotic virial coefficient on one hand and from the composition dependence of $c$ for constant $M$ on the other, has already been reported[3]. This contribution demonstrates that consistent information is also obtained from the analysis of critical demixing data (liquid/liquid equilibria) and from measured vapor pressures (gas/liquid equilibria).

Several aspects of the new concept require further work. Above all it is mandatory to analyze in more detail how the different parameters are made up from enthalpic and entropic contributions. For solvents of marginal quality the experimental material needed for that purpose is comparatively easy to find in the literature in terms of phase diagrams. It is however, much harder to uncover this information for thermodynamically good solvents and over large temperature intervals.